\documentclass[12pt]{article}
\usepackage{graphicx}% Include figure files

\def\beq{\begin{equation}}
\def\eeq{\end{equation}}

\title{%{\it Research Proposal}\\
Light Scattering on Random Dielectric Layers}
\author{O. Fialko and K. Ziegler\\
Institut f\"ur Physik, Universit\"at Augsburg, Germany}
\begin{document}

\maketitle

%\date{\today}
\begin{abstract}
Scattering of light by a random stack of dielectric layers represents a 
one-dimensional
scattering problem, where the scattered field is a three-dimensional
vector field. We investigate the dependence of the scattering properties
(band gaps and Anderson localization) on the wavelength, strength of
randomness and relative angle of the incident wave.
There is a characteristic angular dependence of Anderson localization 
for wavelengths close to the thickness of the layers. In particular, the
localization length varies non-monotonously with the angle. 
In contrast to Anderson localization, absorptive layers do not have this
characteristic angular dependence. 
\end{abstract}

{\bf Keywords:} light scattering, random layers, Anderson localization

%\newpage

\section{Introduction}

Scattering of waves in periodic structures (e.g. in a crystalline
solid material) can be described by Bloch's Theorem
\cite{bloch,kittel87}, a theory
that gives extended, propagating waves. A completely different
situation appears if the periodicity of the scattering structure
is disturbed by disorder: The scattered waves do not propagate any longer
but become {\it localized} due to complex interference effects,
provided that the disorder is sufficiently strong.
This phenomenon, also known as {\it Anderson localization}, was originally
proposed for quantum states \cite{anderson58} and later discussed
in more detail in terms of a renormalization approach by Abrahams et al.
\cite{abraham79}. An important finding of the latter is that the
tendency towards localization is stronger in low dimensions than it is
in higher dimensions. According to this approach, quantum states are
always localized in one and two dimensions, regardless of the strength
of disorder. However, there are exceptions from this result. 
One case, where delocalized states can appear in the presence
of random scattering  in two dimensions, are relativistic (Dirac)
states \cite{ziegler98}. The main difference between nonrelativistic 
(Schr\"odinger) and relativistic (Dirac) states is that the former are 
scalar and the latter are spinor (i.e. vector) states, indicating
that the dimensionality of the state (i.e. scalar vs. vector) plays a
crucial role in the appearance of localization.

Scattering of electromagnetic waves in a random ensemble of scatterers
is a common problem in physics, biology, engineering, and astronomy.
It can be used for a remote analysis of complex objects like, for instance, the
surface of a distant planet. A characteristic feature of light scattering
is coherent backscattering (a comprehensive review can be found 
in \cite{mishchenko06}).
In this context, a more exotic and, at least experimentally, less understood
subject is Anderson localization of light. This has been studied 
in terms of theory and experiment by a number of groups
\cite{john87,raedt89,lag97,scheff99,wiers99,ziegler03,milner05}.
In particular, the existence of Anderson localization for electromagnetic waves
was discussed in Ref. \cite{figotin98}.
In contrast to the localization of quantum states, the electromagnetic field is 
always a three-dimensional vector field. However, there are special situations in 
which the scattering process affects only one component of the electromagnetic vector field.
An example is a wave scattered by a stack of layers with different refractive
indices (cf. Fig. 1): If the wave vector of the incident wave is 
perpendicular to the layers, the vector components are scattered separately, and the 
resulting scattering equation is a scalar one (the Helmholtz equation) 
which is formally equivalent to the Schr\"odinger equation of a quantum state.
The Helmholtz equation has been widely used to study scattering in infinite
and semi-infinite random media 
\cite{akkermans86,Tip91,Klyatskin92,freilikher94,Chang03}. 
For the case of polarization, however, this can only be determined from the vector 
form of the Maxwell equation \cite{stephen86,ozrin92}.

A particular case of a scattering medium is the layered system with a number
of interesting features \cite{zhang95,feng04,hu04,bertolotti05}.
The arrangements of layers is either periodic with an alternating refractive
index or the refractive index changes randomly from layer to layer.
Realizations of layered systems can be found in biological tissues,
in the atmosphere, and in coated optical devices. The effect of Anderson 
localization then can be used to protect and cover an object
if the latter is coated with random layers of non-absorptive materials.
The advantage of this effect is that the electromagnetic waves are prevented to 
enter the coated object without absorbing the electromagnetic waves.

An important aspect of the layered system is that the scattering is
effectively a one-dimensional process. Since the vector components of 
the electromagnetic wave scatter
independently only if the incident wave is perpendicular
to the layers,  by changing the direction
of the incident wave one can couple the components of the electromagnetic wave in
the scattering process. In other words, we can tune our scattering process from
being a scalar one to being a vector one. Thus the scattering by layers
will allow us to study the dependence of localization effects on the scalar
or vector nature of the scattered field.

In this paper we study the localization of electromagnetic waves by
a stack of layers with randomly chosen refractive indices. The strength
of localization is characterized by the Lyapunov exponent of the scattered
wave in the direction perpendicular to the layers. It can be understood as
the inverse of the localization length which can be measured in experiments.
The advantage of considering
the Lyapunov exponent in our calculations is the fact that it is believed to be
a self-averaging quantity \cite{deych98}. This enables us to perform the
calculation without averaging over random ensembles.
Our aim is to compare Anderson localization with absorption, where the latter
is described by an imaginary part in the refractive index \cite{hulst}. Both
effects lead to an exponential decay of the intensity of light along the
scattering process. The difference of their physical origin, however, should
lead to different behavior with respect to wavelength and angle of the
incident light.

\section{The Model}

We consider a 3D system with random dielectric layers that are 
perpendicular to the $x$ direction with
refractive index $n(x)$. For the electric field
${\bf E}(x,y,z)$ we can use the ansatz as a stationary plane wave 
in $y$ and $z$ direction:
\begin{equation}
{\bf E}(x,y,z;t)={\bf E}'(x,k_y,k_z)e^{i(k_y y+k_z z)-i\omega t}.
\end{equation}
Using the short-hand notation ${\bf E}(x)={\bf E}'(x,k_y,k_z)$ and 
$\partial_x\equiv\partial /\partial x$, the electric
field is determined by the Maxwell equation
\beq
\pmatrix{
k_y^2+k_z^2 & ik_y{\partial_x} & ik_z{\partial_x} \cr
ik_y{\partial_x} & k_z^2-{\partial_x}^2 & -k_yk_z \cr
ik_z{\partial_x} & -k_yk_z & k_y^2-{\partial_x}^2 
}{\bf E}(x)=\epsilon(x){\bf E}(x)\ ,\ \ \  \epsilon(x)=n(x)^2(\omega/c)^2.
\label{maxwell}
\eeq
For discrete layers and not too short wavelengths $\lambda=2\pi c/\omega$
(i.e. for a wavelength $\lambda$ larger than the thickness of the layers), 
Eq. (\ref{maxwell}) can be written in a discrete form
by integrating the Maxwell equation in $x$ direction from a layer boundary $x_n$
to the next layer boundary $x_{n+1}$. Since ${\bf E}(x)$ does not change much 
within a layer of  thickness $d_n=x_{n+1}-x_n$ if $\lambda>d_n$,
we can replace its value for $x_n\le x<x_{n+1}$ by 
\begin{equation}
{\bf E}_n=\frac{1}{d_n}\int_{x_n}^{x_{n+1}}{\bf E}(x)dx
\approx{\bf E}(x_n) \ .
\end{equation}
Moreover, the differential operator 
$\partial_x$ is replaced by a difference operator as
\begin{equation}
\partial_x{\bf E}(x)\to\frac{1}{d_n} \int_{x_n}^{x_{n+1}}\partial_x{\bf E}(x)dx=
\frac{{\bf E}_{n+1}-{\bf E}_{n-1}}{d_n} \ .
\end{equation}
In other words, $x$ is replaced by the discrete coordinate $n$.
For the following study we assume that the layers have the same thickness
(but varying refractive index) and we choose this to be $d_n=d$. Then 
all length scales are given in units of the thickness of the layers.
Thus Eq. (\ref{maxwell}) can be written as
\beq
A_2({\bf E}_{n+1}+{\bf E}_{n-1}-2{\bf E}_{n})+A_1({\bf E}_{n+1}-{\bf E}_{n-1})
+A_0{\bf E}_{n}=\epsilon_n{\bf E}_{n}
\eeq
with
\[
A_2=\pmatrix{
0 & 0 & 0 \cr
0 & -1 & 0 \cr
0 & 0 & -1 \cr
},\ \ \ 
A_1={1\over2}\pmatrix{
0 & ik_y & ik_z \cr
ik_y & 0 & 0 \cr
ik_z & 0 & 0 \cr
},
\]
\begin{equation}
A_0=\pmatrix{
k_y^2+k_z^2 & 0 & 0 \cr
0 & k_z^2 & -k_yk_z \cr
0 & -k_yk_z & k_y^2 \cr
}.
\end{equation}
This can also be expressed as a recursion relation of ${\bf E}_{n}$:
\beq
C^{-1}_nA{\bf E}_{n+1}+C^{-1}_nB{\bf E}_{n-1}={\bf E}_{n}
\label{recursion}
\eeq
with
\begin{equation}
A=A_2+A_1,\ \ \ B=A_2-A_1,\ \ \ C_n=-A_0+2A_2+\epsilon_n{\bf 1}.
\end{equation}
%\subsection{Two-component Representation}
Multiplication of Eq. (\ref{recursion}) by $C_n$ gives
\begin{equation}
A{\bf E}_{n+1}=C_n{\bf E}_{n}-B{\bf E}_{n-1}
\end{equation}
and a subsequent shift of $n$ by 1 gives
\begin{equation}
A{\bf E}_{n+2}-C_{n+1}{\bf E}_{n+1}=-B{\bf E}_{n}.
\end{equation}
These two equations can be combined to a first order difference
equation as
\begin{equation}
\pmatrix{
A & 0 \cr
-C_{n+1} & A \cr
}
\pmatrix{
{\bf E}_{n+1} \cr
{\bf E}_{n+2} \cr
}
=\pmatrix{
-B & C_n \cr
0 & -B \cr
}
\pmatrix{
{\bf E}_{n-1} \cr
{\bf E}_{n} \cr
}.
\end{equation}
Now we multiply this equation from the left with
\begin{equation}
\pmatrix{
A & 0 \cr
-C_{n+1} & A \cr
}^{-1}
=\pmatrix{
A^{-1} & 0 \cr
A^{-1}C_{n+1}A^{-1} & A^{-1} \cr
}
\end{equation}
and introduce the new (six-dimensional) vector field
\begin{equation}
{\bf \Psi}_n=
\pmatrix{
{\bf E}_{n-1} \cr
{\bf E}_{n} \cr
}
\end{equation}
to write 
\beq
{\bf \Psi}_{n+2}={\hat T}_n{\bf \Psi}_{n}
\label{recursion2}
\eeq
with the transfer matrix
\[
{\hat T}_n=
\pmatrix{
A^{-1} & 0 \cr
A^{-1}C_{n+1}A^{-1} & A^{-1} \cr
}
\pmatrix{
-B & C_n \cr
0 & -B \cr
}
\]
\beq
=\pmatrix{
-A^{-1}B & A^{-1}C_n \cr
-A^{-1}C_{n+1}A^{-1}B & A^{-1}C_{n+1}A^{-1}C_n-A^{-1}B \cr
}.
\label{transfer}
\eeq
The transfer matrix ${\hat T}_n$ satisfies $\det({\hat T}_n)=1$.
Eq. (\ref{recursion2}) will be used as the starting
point for our subsequent calculations. In particular, we can iterate this equation
and obtain
\beq
{\bf \Psi}_{2n+2}={\hat T}_{2n}\cdots{\hat T}_2{\hat T}_0{\bf \Psi}_{0}.
\label{solution}
\eeq

The special case of a plane wave propagating only in $x$ direction
(i.e. $k_y=k_z=0$) leads to scalar equations for
the electric field, since the Maxwell equation
(\ref{maxwell}) decomposes into two independent scalar equations 
for $E_y$ and $E_z$. For $k_y=k_z=0$ the transfer matrix becomes
a $2\times 2$ matrix, where the $3\times3$ matrices $A$ and $B$
are identical and equal to the scalar $-1$. 
Moreover, $C_n$ also becomes a scalar: $-2+\epsilon_n$. As a
consequence, the transfer matrix is
\beq
{\hat T}_n=\pmatrix{
-1 & -h_n \cr
h_{n+1} & h_{n+1}h_n-1 \cr
},\ \ \ h_n=\epsilon_n-2 .
\label{transfer0}
\eeq 

\section{Results}

Elastic scattering of waves can be characterized by various physical
quantities. A very useful quantity to characterize localization
effects in our one-dimensional scattering
geometry of the stacked layers is the Lyapunov exponent
\cite{deych98,lifshitz88}:
\beq
\gamma_j=\lim_{n\to\infty}{1\over n}\log(|E_{n,j}|)\ \ \ \ (j=x,y,z).
\label{lyapunov}
\eeq
It measures the exponential decay of the magnitude of the wave due to
scattering in the random medium. A non-localized wave has $\gamma_j=0$,
whereas $\gamma_j\ne 0$ describes a localized wave $E_{n,j}$. The larger
$|\gamma_j|$ the stronger the localization is. 
The self-averaging property of the Lyapunov exponent \cite{deych98} is a crucial
advantage for our numerical calculations: there is no need for averaging
over an ensemble of random scatterers because this is achieved by choosing 
a sufficiently large stack of randomly chosen layers.  

\subsection{Analytic Results: Alternating Layers}

In the case of layers with alternating value of $\epsilon_n$
\begin{equation}
\epsilon_n=\cases{
\epsilon_e=n_e^2\omega^2/c^2 & for $n$ even \cr
\epsilon_o=n_o^2\omega^2/c^2 & for $n$ odd \cr
} \ ,
\end{equation}
the problem in Eq. (\ref{recursion2}) is translational invariant:
${\hat T}_n={\hat T}$. Therefore, it can be solved easily by diagonalizing
the $6\times6$ matrix ${\hat T}$. For $k_y=k_z=0$ (scalar case)
we have only the two parameters $h_e, h_o$ in the transfer matrix of Eq. (\ref{transfer0}).
Moreover, in Eq. (\ref{solution}) we need only the $2\times2$ version
${\hat T}_n$  of Eq. (\ref{transfer0}) for even $n$:
\begin{equation}
{\hat T}_{2k}=\pmatrix{
-1 & -h_e \cr
h_o & h_oh_e-1 \cr
}\ \ \ \ (h_{e,o}=\epsilon_{e,o}-2).
\end{equation}
Thus the corresponding eigenvalues are
\begin{equation}
\lambda_\pm=-1+h_oh_e/2\pm\sqrt{-h_oh_e+h_o^2h_e^2/4}.
\end{equation}
The eigenvalues are complex for $0<h_oh_e<4$ with $|\lambda_\pm|=1$.
They represent propagating solutions
\begin{equation}
{\bf \Psi}_{2n}=\pmatrix{
(\lambda_+)^n & 0 \cr
0 & (\lambda_-)^n \cr
}{\bf \Psi}_{0}
\equiv\pmatrix{
e^{i\kappa n} & 0 \cr
0 & e^{-i\kappa n} \cr
}{\bf \Psi}_{0}
\end{equation}
with the real ``wave vector'' $\kappa$ which satisfies
\beq
\cos \kappa={h_oh_e\over2}-1.
\label{gap}
\eeq
Solving this equation gives a dispersion $\omega^2(\kappa)$ 
(cf. Fig. 2) with a gap $\Delta$, opening at $\kappa=\pm\pi$:
\begin{equation}
\Delta=2\left| {1\over n_e^2}-{1\over n_0^2}\right| .
\end{equation}

For $h_oh_e>4$ we have pairs of real eigenvalues with $\lambda_+\lambda_-=1$
and no propagating solution. The six eigenvalues $\lambda_j$ (only
$|\lambda_j|$ is plotted) of the vector case (i.e. for $k_x=k_y\ne0$) 
are shown in Figs. 3-6. Propagating solutions (i.e. $|\lambda_j|=1$) are
found for different regimes of $\omega^2/c^2$. 

\subsection{Numerical Results}

In the case of scattering by random layers we rely on a numerical 
procedure. Performing such a calculation is relatively easy because 
we only need to multiply $6\times 6$ (or $2\times2$ in the scalar 
case $k_y=k_z=0$) randomly chosen transfer matrices. Then the Lyapunov exponent
can be calculated from the product according to Eq. (\ref{lyapunov}).
However, when we perform many multiplications of transfer matrices, 
numerical accuracy plays a crucial role. The numerics is dominated
by large eigenvalues but we are interested in eigenvalues 
$\lambda_j$ with $|\lambda_j|\approx 1$. The accuracy of the latter
is suppressed by the large eigenvalues. In order to avoid this
problem we apply a method which orthonormalizes the columns of the 
product matrix after a few multiplications via the Gram-Schmidt procedure. 
This process separates automatically the different exponentially growing
contributions \cite{pichard81,mackinnon83}. Then the logarithms of the lengths 
of the vectors are stored. The Lyapunov exponents are given by the mean value 
of these logarithms divided by the number of steps between orthonormalizations. 
Finally, the smallest Lyapunov exponent is stored. It was found for one-dimensional 
systems that the number of multiplications 
required for convergence is inversely proportional to the Lyapunov exponent and 
it is approximately given by 
\cite{mackinnon83}
\begin{equation}
N_{\rm max}\approx 2(\epsilon^2\gamma)^{-1},
\end{equation}  
where $\epsilon$ is the relative accuracy. In order to reach convergency of 
$\epsilon\sim 0.01$ we calculate typically up to $10^6$ layers (cf. Fig. 7).

\section{Discussion and Conclusions}

A stack of layers with alternating refractive indices $n_e$ and $n_o$
presents an instructive example for the influence of the wave vector
of the incident wave on the scattering properties. The relative angle $\varphi$
of the incident wave with the layers is given through
\beq
\cos\varphi= \sqrt{1-c^2(k_y^2+k_z^2)/\omega^2} .
\eeq
If we start with the scalar case (i.e. $k_y=k_z=0$ or $\varphi=0$)
we find two bands of propagating waves with a gap at the 
boundaries of the Brillouin zone (cf. Fig. \ref{fig2}). As soon as we
introduce a small nonzero wave vector parallel to the
layers (i.e. $k_y=k_z\ne 0$ or $\cos\varphi<1$), the transfer matrix 
becomes six-dimensional. The gap between the
two bands of propagating solutions remains almost unaffected
(cf. Fig. \ref{fig3}). 
Besides the two eigenvalues of the scalar case, related to the 
propagating solution, there are also eigenvalues $|\lambda_j|\ne1$ 
which are not related to solutions of our scattering problem.

For $\varphi>0$ the band of propagating solutions
(i.e. for eigenvalues $|\lambda_j|=1$) persists, together with the band gap
(cf. Fig. \ref{fig5} for $k_y=k_z=2$) but the band gap disappears for 
$n_o=n_e$ (cf. Fig. \ref{fig4}), as we anticipate from the scalar case.
However, in contrast to the scalar case, there is only a tiny band gap
for $n_o\ne n_e$, as shown in Fig. \ref{fig6}.

In the case of random layers we consider randomly independent fluctuations of
the refractive index in Eq. (\ref{maxwell}) 
\beq
n_n^2=1 + \alpha h_n \ ,
\eeq
where $h_n$ is a random variable, distributed uniformly on the
interval $[-0.5,0.5]$. The parameter $\alpha$ controls the strength of the
random fluctuations. We find that (Anderson) localization is increasingly 
efficient with decreasing wavelength $\lambda=2\pi c/\omega=2\pi/k$.
This is plausible, since a wave with large wavelength experiences the randomly fluctuating
refractive index as an average refractive index. 
In Fig. \ref{fig8} the Lyapunov exponent $\gamma_x$ is plotted as a function of $\alpha$,
where $\gamma_x$ increases monotonously with $\alpha$ and $k$. 
On the other hand, the behavior of the Lyapunov exponent as a function of the
angle $\varphi$ is not so obvious. For $\omega/c\approx 1$ it has a non-monotonous
behavior (cf. Fig. \ref{fig9}): the Lyapunov exponent decreases linearly with
$\cos\varphi$, reaches a minimum near $\cos\varphi=0.8$ and then increases.
For $\omega/c\approx 1.2$ there is well-developed minimum on the interval 
$0.6<\cos\varphi<0.8$ , as shown in Fig. \ref{fig11}. 
For larger values of $\omega/c$ (i.e. shorter wavelengths) the Lyapunov exponent
is almost constant as a function of $\varphi$. The non-monotonous behavior of
$\gamma_x$ with respect to the angle $\varphi$ might be related to the fact that
the mechanism of localization is due to interference of different parts of the
scattered wave. The distance between two sucessive scattering events, which 
happen at the interfaces of the layers, depends of the angle $\varphi$.
Changing $\varphi$ implies a change of the distance between the scattering events and, 
therefore, allows us to change between constructive and destructive interference. 
  
Absorption is another mechanism which leads to an exponential decay of 
the wave $E_{n,j}$. It can be included in our Maxwell equation
(\ref{maxwell}) by adding an imaginary part to the refractive index \cite{hulst}:
\beq
n^2_n +i\eta \ .
\eeq
In a homogeneous medium absorption should not be very
sensitive to a change of the angle $\varphi$. 
Assuming identical absorptive layers (i.e. $\alpha=0$) we find, in contrast to the case of
random layers, no characteristic variation of the Lyapunov exponent (cf. Fig. \ref{fig10}).
This effect allows us to distinguish Anderson localization and absorption in a scattering
experiment by measuring the variation of the Lyapunov exponent as a function of the
angle $\varphi$.  

In conclusion, we have studied the propagation of an electric field in a stack of equally
thick layers. Layers with alternating refractive index do not allow propagation for
some wavelengths by opening a band gap. Layers with randomly changing refractive index, 
on the other hand, have no propagating solution but experience Anderson localization.
The strength of the latter depends crucially on the wavelength and the relative angle
of the incident wave. In contrast to Anderson localization, absorption does not
show a characteristic dependence on this angle.

\newpage

\begin{figure} 
\begin{center}
\includegraphics[width=12cm]{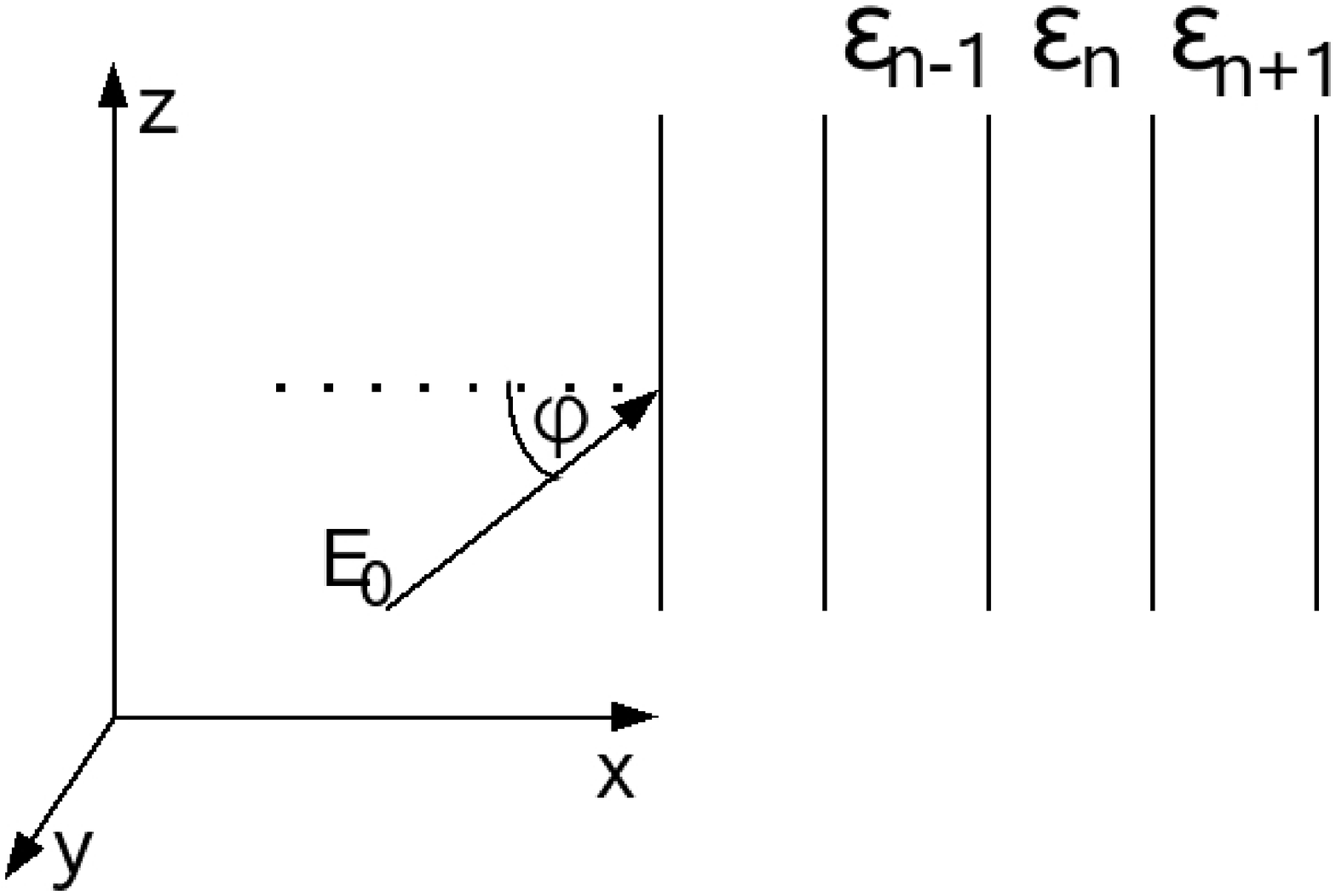}
\end{center} 
\caption{
Scattering of an electromagnetic wave on layers
with refractive index $\epsilon_n$.
}
\label{fig1}
\end{figure}

\begin{figure} 
\begin{center}
\includegraphics[width=12cm]{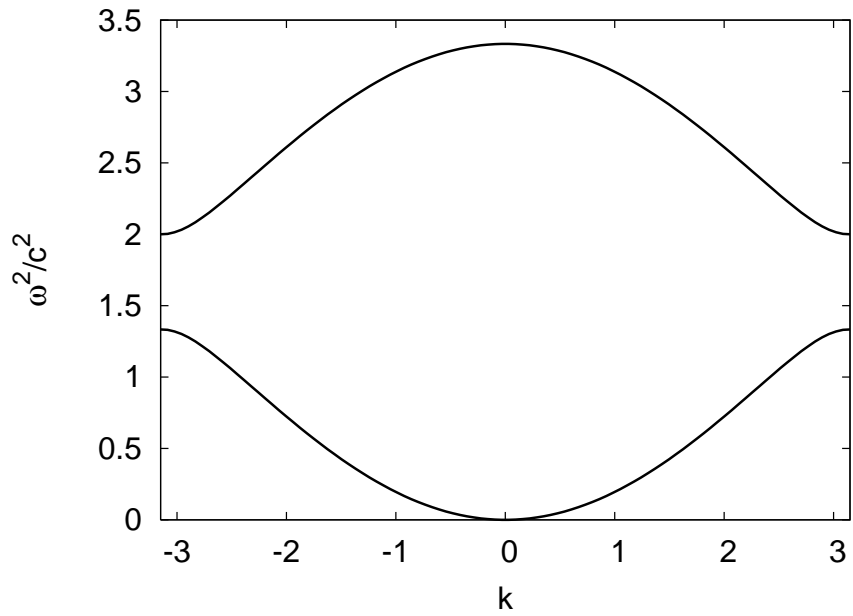}
\end{center} 
\caption{Dispersion $(\omega/c)^2(\kappa)$ of alternating layers 
with refractive indices $n_e=1.000,n_o=1.025$ and $k_y=k_z=0$ (scalar case). 
There is a gap opening at the edges of the Brillouin zone.}
\label{fig2}
\end{figure}

\begin{figure} 
\begin{center}
\includegraphics[width=12cm]{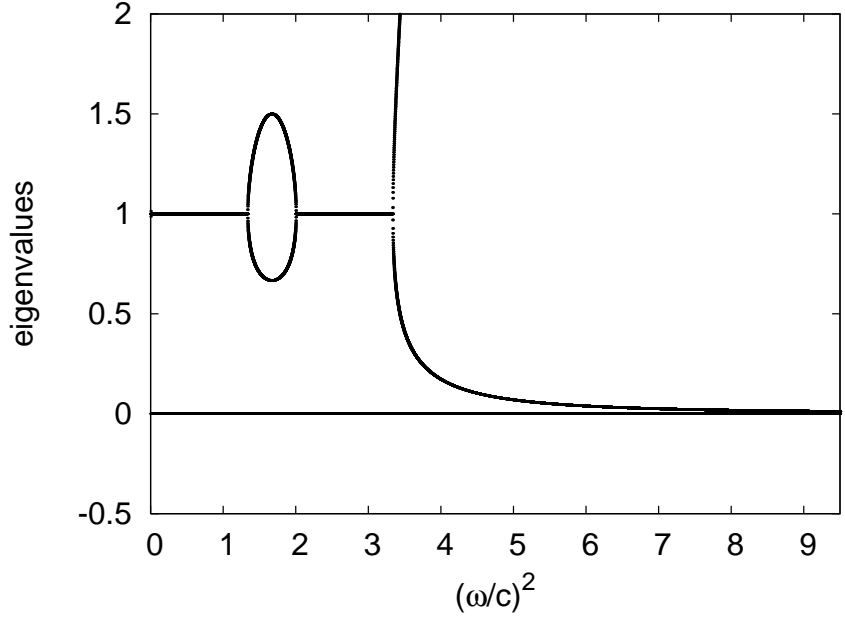}
\end{center} 
\caption{
Eigenvalues of the $6\times 6$ transfer matrix for an alternating
array of layers with the refractive indices of Fig. 2
and $k_y=k_z=0.001$ (close to the scalar case). The gap is
not affected by the very small change of $k_y=k_z$ in
comparison with Fig. 2. 
}
\label{fig3}
\end{figure}

\begin{figure} 
\begin{center}
\includegraphics[width=12cm]{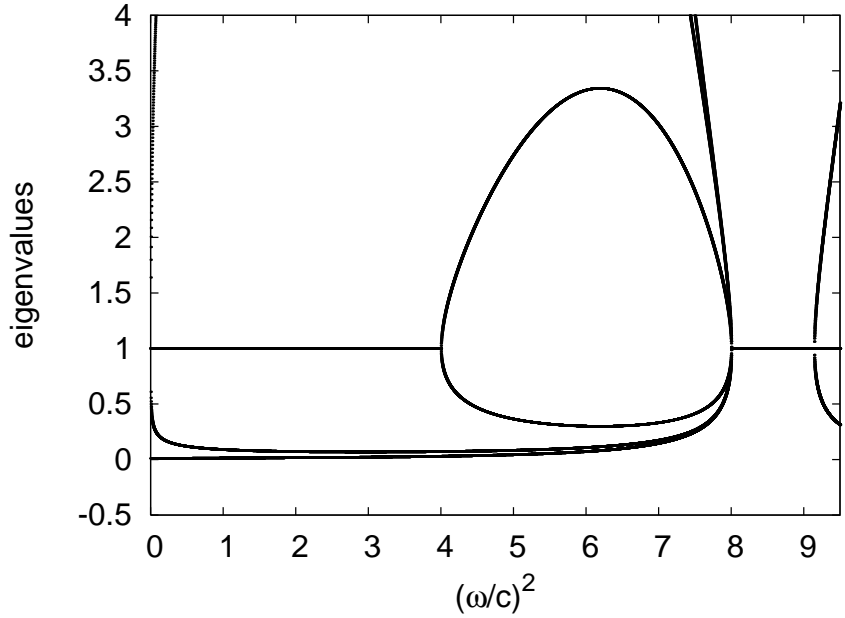}
\end{center} 
\caption{The same as Fig. 3 but for $n_e=n_o=1$ (homogeneous case)
and $k_y=k_z=2$.
}
\label{fig4}
\end{figure}

\begin{figure} 
\begin{center}
\includegraphics[width=12cm]{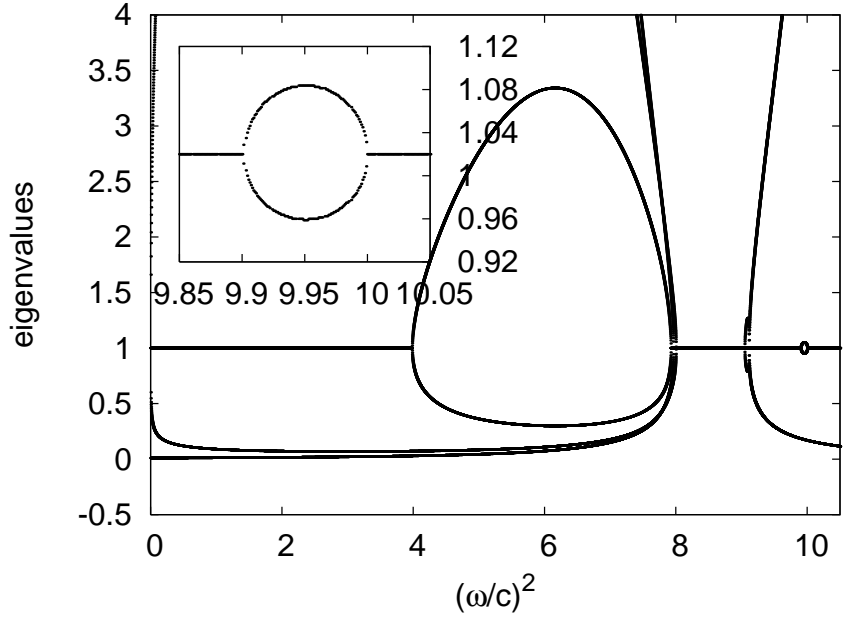}
\end{center} 
\caption{The same as Fig. 4 but for %$n^2_e = 1.0, n^2_o = 1.01$
$n_e = 1.000, n_o = 1.005$
and $k_y=k_z=2$. A propagating solution
has a gap between $\omega^2\approx 9.9c^2$ and $10.0c^2$ (see inset). 
}
\label{fig5}
\end{figure}

\begin{figure} 
\begin{center}
\includegraphics[width=12cm]{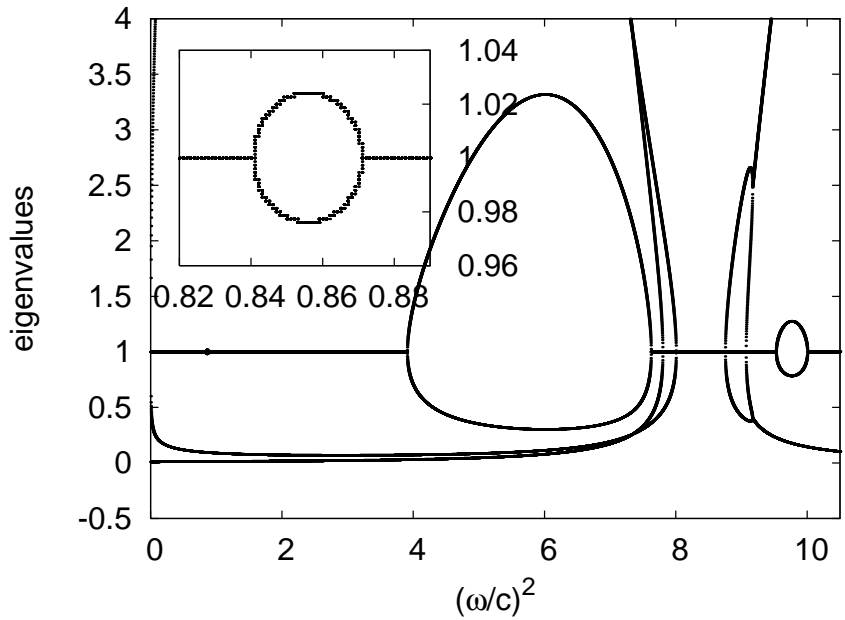}
\end{center} 
\caption{The same as Fig. 5 but for %$n^2_e = 1.0, n^2_o = 1.05$.
$n_e = 1.000, n_o = 1.025$.
A new gap for the band of the propagating solution develops near $\omega^2\approx 0.85 c^2 $.
}
\label{fig6}
\end{figure}

\begin{figure} 
\begin{center}
\includegraphics[width=12cm]{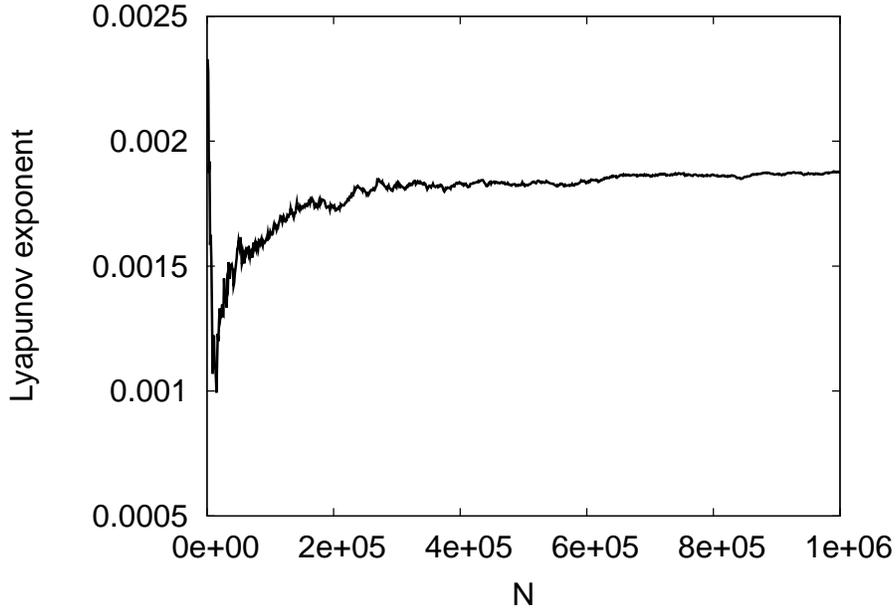}
\end{center} 
\caption{
Scattering by random layers:
Lyapunov exponent for $E_x$ as a function of the number of random 
layers $N$ for $k_z=k_y=1.5$ with $n_n^2(\omega/c)^2=\epsilon_n = 5 + 0.2 h_n$.
$h_n$ is a random variable with $-0.5\le h_n\le 0.5$.
}
\label{fig7}
\end{figure}

\begin{figure} 
\begin{center}
\includegraphics[width=12cm]{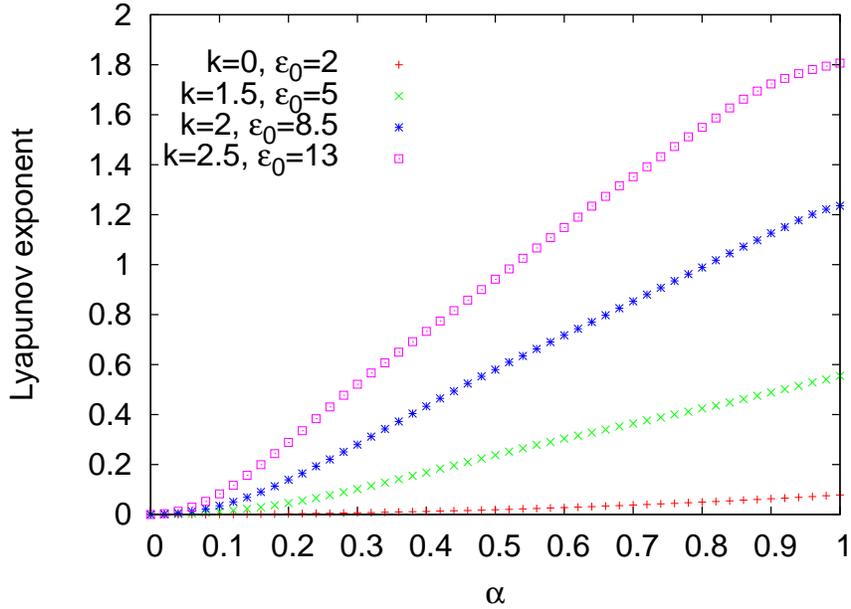}
\end{center} 
\caption{Behavior of the Lyapunov exponents for various wave vectors $k=k_y=k_z$ and 
$\epsilon_n = \epsilon_0 (1 + \alpha h_n)$.
$h_n$ is the random variable of the previous figure.
%Initial conditions are $E_x=E_y=E_z=1.0$ for zeroth and first layers.
$\epsilon_0 = 2$ (for $k=0$), 
$\epsilon_0 = 5$ (for $k=1.5$), 
$\epsilon_0 = 8.5$ (for $k=2$), 
$\epsilon_0 = 13$ (for $k=2.5$).
%$E_x, E_y,E_z$ for n-th layer.
The wavelength in the direction perpendicular to the layers is $2\pi/k$, in units of the thickness
of the layers. 
}
\label{fig8}
\end{figure}

\begin{figure}
\begin{center}
\includegraphics[width=14cm]{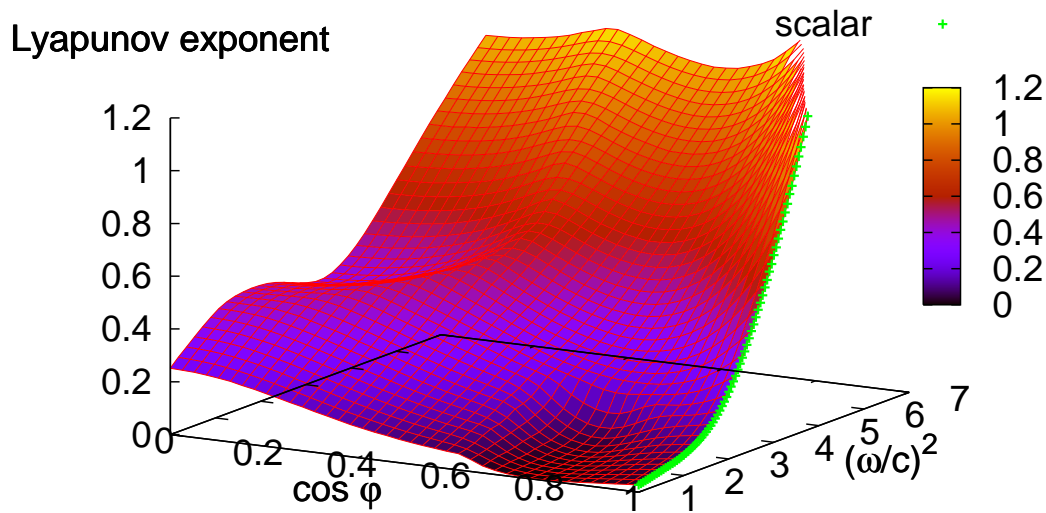}
\end{center}
\caption{Behavior of the Lyapunov exponents for random layers with $\alpha=1$ 
as a function of the angle $\varphi$ of the incident wave and $\omega^2/c^2$.}
\label{fig9}
\end{figure}

\begin{figure}
\begin{center}
\includegraphics[width=14cm]{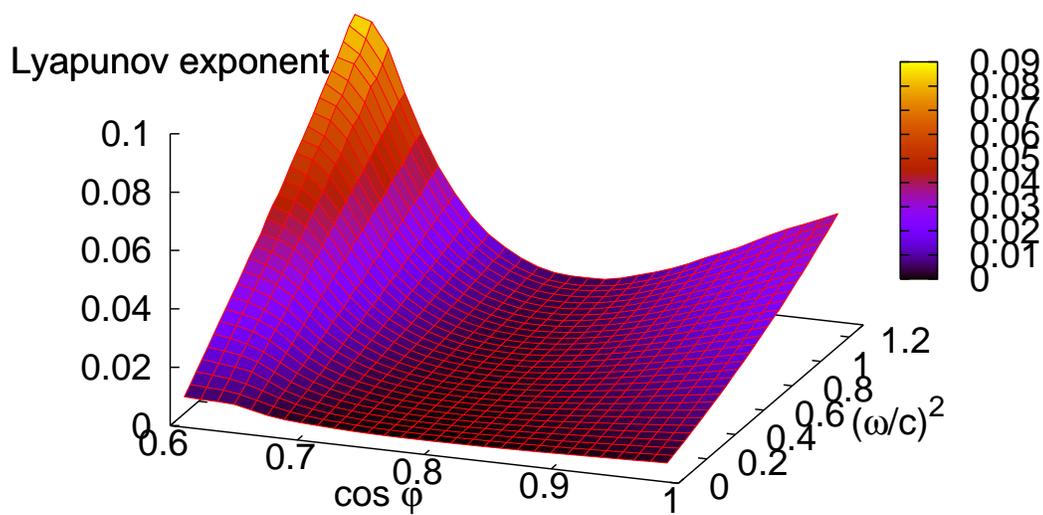}
\end{center}
\caption{Behavior of the Lyapunov exponents for a system of random layers $\alpha=1$.}
\label{fig11}
\end{figure}

\begin{figure}
\begin{center}
\includegraphics[width=14cm]{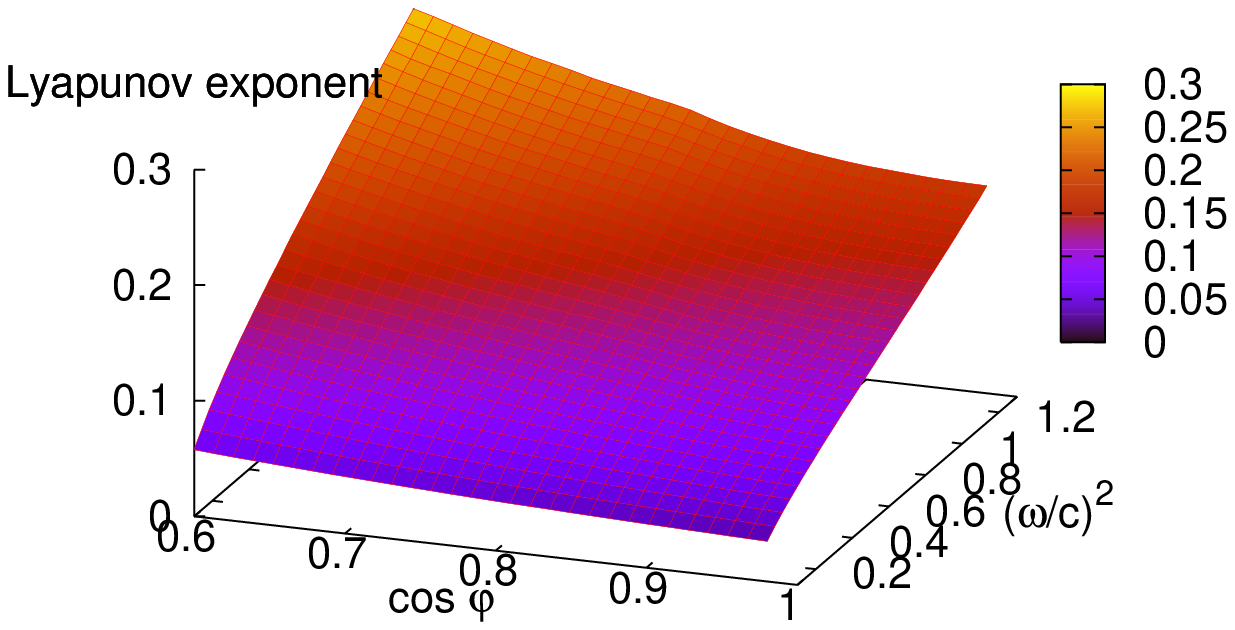}
\end{center}
\caption{Behavior of the Lyapunov exponents for absorbing layers with $\eta=0.1$}
\label{fig10}
\end{figure}


\begin{thebibliography}{}

\bibitem[1]{bloch}
Bloch F. Z Physik 1928:{\bf 52}:555-600.
%http://en.wikipedia.org/wiki/Bloch's\_Theorem
%http://planetmath.org/encyclopedia/BlochsTheorem.html

\bibitem[2]{kittel87}
Kittel C. Quantum Theory of Solids. New York: J. Wiley, 1987.

\bibitem[3]{anderson58}
Anderson PW. Phys Rev 1958:109:1492-1505.

\bibitem[4]{abraham79}
Abrahams E, Anderson PW, Licciardello DC, Ramakrishnan TV.
Phys Rev Lett 1979:42:673-676.

\bibitem[5]{ziegler98}
Ziegler K. Phys Rev Lett 1998:80:3113-3116.

\bibitem[6]{mishchenko06}
Mishchenko MI, Travis LD, Lacis AA.
Multiple Scattering of Light by Particles: Cambridge: Cambridge University Press, 2006.

\bibitem[7]{john87}
John S. Phys Rev Lett 1987:58:2486-2489.

\bibitem[8]{raedt89}
De Raedt H, Lagendijk A, de Vries P.
Phys Rev Lett 1989:62:47-50.

\bibitem[9]{lag97}
Wiersma DS, Bartolini P, Lagendijk A, Righini R.
Nature 1997:{\bf 390}:671-673.

\bibitem[10]{scheff99}
Scheffold F, Lenke R, Tweer R, Maret G.
Nature 1999:398:206-207.

\bibitem[11]{wiers99}
Wiersma DS, Rivas JG, Bartolini P, Lagendijk A, Righini R.
Nature 1999:398:207.

\bibitem[12]{ziegler03}
Ziegler K. J Quant Spec Rad Transf 2003:79-80:1189-1198.

\bibitem[13]{milner05}
Milner V, Genack AZ. Phys Rev Lett 2005:94:073901-1-4.

\bibitem[14]{figotin98}
Figotin A, Klein A.
J Opt Soc Am A 1998:15:1423-1435.

\bibitem[15]{akkermans86}
Akkermans E, Wolf PE, Maynard R.
Phys Rev Lett 1986:56:1471-1474.

\bibitem[16]{Tip91}
van Tiggelen BA, and Tip A. 
J Phys I France 1991:1:1145-1154

\bibitem[17]{Klyatskin92}
Klyatskin VI, and Saichev AI. Uspechi Phys Nauk 1992:162:161-194

\bibitem[18]{freilikher94}
Freilikher V, Pustilnik M, Yurkevich I.
Phys Rev Lett 1994:73:810-813.

\bibitem[19]{Chang03}
Chang SH, Cao H, and Ho ST. 
IEEE Journal of quantum electronics 2003:39:364-374.

\bibitem[20]{stephen86}
Stephen MJ, Cwilich G.
Phys Rev B 1986:34:7564-7572.

\bibitem[21]{ozrin92}
Ozrin VD.
Phys Lett A 1992:162:341-345.

\bibitem[22]{zhang95}
Zhang Z-Q.
Phys Rev 1995:B52:7960-7964.

\bibitem[23]{feng04}
Feng Y, Ueda K.
Optics Express 2004:12:3307-3311.

\bibitem[24]{hu04}
Hu L, Schmidt A, Narayanaswamy A, Chen G.
Journal of Heat Transfer 2004:126:786-792.

\bibitem[25]{bertolotti05}
Bertolotti J, Stefano G, Wiersma DS, Ghulinyan M, Pavesi L.
Phys Rev Lett 2005:94:113903-1-4.

\bibitem[26]{deych98}
Deych LI, Zaslavsky D, and Lisyansky AA. 
Phys Rev Lett 1998:81:5390-5393.

\bibitem[27]{hulst}
van de Hulst HC. Light Scattering by Small Particles. New York: Dover
Publications, 1981. 

\bibitem[28]{lifshitz88}
Lifshitz IM, Gredeskul SA, and Pastur LA. Introduction to the
Theory of Disordered Systems. New York: J. Wiley, 1988.

\bibitem[29]{pichard81}
Pichard JL, and Sarma G. J. Phys. C: Solid State Physics 1981:14:L127-L132.

\bibitem[30]{mackinnon83}
MacKinnon A, and Kramer B.  Z Physik 1983:B53:1-13.


\end{thebibliography}
\end{document}